\documentclass[useAMS,usenatbib]{mn2e}

\usepackage{epsfig}

\title[DY\,Centauri's nebula]{The changing nebula around the hot R Coronae Borealis star DY Centauri}

\author[N. Kameswara Rao et al.]{N. Kameswara Rao$^{1,2}$, David L. Lambert$^{2}$, D. A. Garc\'{\i}a-Hern\'{a}ndez$^{3,4}$,\and \& Arturo Manchado$^{3,4,5}$\\
          $^1$ 543, 17$^{th}$ Main, IV Sector, HSR Layout, Bangalore 560102, India \\
          $^2$The W.J. McDonald Observatory, University of Texas, Austin, TX 78712-1083, USA\\
          $^3$Instituto De Astrof\'{\i}sica De Canarias, V\'{\i}a L\'actea s/n, E-38200 La Laguna, Tenerife, Spain\\
          $^4$Departmento de Astrof\'{\i}sica, Universidad de La Laguna(ULL), E-38206 La Laguna, Tenerife, Spain \\
          $^5$Consejo Superior de Investigaciones Cient\'{\i}ficas (CSIC), Spain}
\begin{document}
\date{Accepted
         Received ;
         in original form  }

\pagerange{\pageref{firstpage}--\pageref{lastpage}}
\pubyear{}

\maketitle

\label{firstpage}

\begin{abstract}

Among the distinguishing characteristics of the remarkable hot R Coronae
Borealis star DY Cen, which was recently found to be a spectroscopic binary, is
the presence of nebular forbidden lines in its optical spectrum. A compilation
of photometry from 1970 to the present suggests that the star has evolved to
higher effective temperatures. Comparison of spectra from 2010 with earlier
spectra show that between 2003 and 2010, the 6717 and 6730 \AA\ emission lines
of [S\,{\sc ii}] underwent a dramatic change in their fluxes suggesting an
increase in the nebula's electron density of 290 cm$^{-3}$  to  3140 cm$^{-3}$
from 1989 to 2010 while the stellar temperature increased from 19500 K to 25000
K. The  nebular radius is about 0.02 pc, 60000 times bigger than the
semimajor axis of DY Cen binary system. Rapid changes of stellar temperature
and its response by the nebula demonstrate stellar evolution in action.

\end{abstract}


\section{Introduction}

DY Centauri is  known as a hot  R Coronae Borealis (RCB) star. RCB stars are a
rare class of peculiar variable stars with two principal  defining
characteristics: (i) they exhibit a propensity to fade at unpredictable times by
up to about eight magnitudes as a result of obscuration by clouds of soot, and
(ii) they have a supergiant-like atmosphere that is very H-deficient, He-rich
and C-rich. The subject of this paper, the remarkable star DY Cen, one of the
hottest RCBs, is even a peculiar RCB member on several accounts. It is one of
the most hydrogen-rich RCBs. Its circumstellar environment may be home to the
fullerene C$_{60}$ or more likely proto-fullerenes (Garc\'{\i}a-Hern\'{a}ndez et
al. 2011a; 2012). In terms of its chemical composition, DY Cen may have a
composition that sets it apart from most RCB and extreme helium (EHe) stars --
see Jeffery \& Heber's (1993) abundance analysis: DY Cen is not only
uncharacteristically H-rich but is very Fe-poor with a very high S/Fe ratio,
thus has been classified as a minority RCB star (Lambert \& Rao 1994);
i.e., a RCB with extraordinarily high Si/Fe and S/Fe ratios. However, Jeffery
et al. (2011) suggest that the Fe abundance was greatly underestimated in 1993.
Finally, DY Cen is a single-lined spectroscopic binary (Rao et al. 2012), the
only known binary among RCBs.

The origins of the hydrogen-deficient RCB and EHe stars are not presently fully
understood. Two proposals  are commonly advocated. In one, a final helium shell
flash or a very late thermal pulse occurs on a cooling white dwarf star (Iben et
al. 1983; Renzini 1990; Herwig 2000) which  swells the envelope of  the star to
supergiant dimensions for a few thousand years and the remaining hydrogen is
convected in and consumed while  helium and carbon are convected out to the
surface. Living examples are believed to be Sakurai's object (V4334 Sgr) and FG
Sge (Asplund et al. 1997; Jeffery \& Sch\"onberner 2006). The other proposal
involves the merger of two white dwarfs: a carbon-oxygen white dwarf accretes a
helium white dwarf as a close binary orbit shrinks under the influence of energy
loss by gravitational radiation (Webbink 1984; Iben \& Tutukov 1986). The merger
leads to a swollen envelope around the C-O white dwarf which lasts a few
thousand years. Evidence from the chemical compositions of RCB and EHe stars
suggests that most are products of a merger (Garc\'{\i}a-Hern\'{a}ndez et al.
2009, 2010; Pandey \& Lambert 2011; Jeffery et al. 2011).

DY Cen with an orbital period of about 39 days is experiencing  mass loss and
presumably mass transfer to a low mass companion.\footnote{A merger of the two
stars through loss of gravitational energy will not occur for about 10$^{15}$
years.} Moreover, it is likely to have experienced mass loss and transfer
previously in arriving at its H-poor condition. Rao et al. (2012) suggest DY Cen
may evolve to a He-rich sdB star. Many such sdB stars are binaries. In this
sense, DY Cen is a representative of a third way to form a RCB star. The fact
that its supergiant phase is short-lived relative to life as a sdB accounts for
the rarity of such hot RCBs relative to the sdB population.

The presence of forbidden emission lines in DY Cen's optical spectrum, was
noticed in 1989 (Rao et al. 1993). These lines were attributed to a nebula
around DY Cen. In this paper, we analyse the nebular lines  and their evolution
over two decades. Nebular lines were especially prominent in a 2010
high-resolution spectrum with broad wavelength coverage. Inspection of earlier
spectra show that physical conditions in the region emitting the nebular lines
have changed since 1989. In particular, the electron density inferred from the 
ratio of the [S\,{\sc ii}] 6717 and 6731 \AA\ lines increased about eight-fold
from 1989 to 2010.

\section{Spectroscopic Observations}

The general properties of DY Cen's spectrum have been described by Pollacco \&
Hill (1991), Jeffery \& Heber (1993), Rao, Giridhar \& Lambert (1993), Giridhar,
Rao \& Lambert (1996) and De Marco et al. (2002).  Optical  spectra are 
presently a combination of photospheric absorption lines (e.g. O\,{\sc ii},
N\,{\sc ii}, Si\,{\sc iii} lines), emission lines due to a stellar wind  (mainly
C\,{\sc ii}, He\,{\sc i} often superposed on underlying absorption lines),  and
nebular emission lines of [S\,{\sc ii}], [N\,{\sc ii}], [O\,{\sc i}], [Fe\,{\sc
ii}], etc. In addition, circumstellar and interstellar lines  from Ca\,{\sc ii},
Na\,{\sc i}, K\,{\sc i} and other species as well as the enigmatic  diffuse
interstellar bands (DIBs) are present in absorption (Garc\'{\i}a-Hern\'{a}ndez
et al. 2012).

This paper considers new spectra from 2010 acquired as a result of an ESO
Director's Discretionary Proposal. The spectra were obtained on four nights in
February-March 2010 with the cross-dispersed echelle spectrograph UVES at the
VLT at ESO's Paranal Observatory. The UVES spectrograph was used with the
1.2" slit (slit length of 8" and P.A.=0 degrees) and the standard setting DIC2
(390+760). The seeing was always better than 1.5" and we obtained 11 individual
exposures of 1800 s each (total exposure time of 5.5 hours). Since the nebular
lines are not expected to change in a short time, the individual exposures were
combined to increase the signal-to-noise (S/N) ratio. The S/N in the continuum
at 4000 \AA~is $\sim$200 and at wavelengths longer than 6000 it is higher than
250. The slit width is just about the size of the nebular diameter, as estimated
later. The spectral resolving power is about 37000, as estimated from O$_2$
telluric lines at 6970 \AA. Wavelength coverage is from 3300 to 4500 \AA, 5700
to 7525 \AA\ and 7660 to 9460 \AA. Along with DY Cen, HD 115842 a nearby B
supergiant was observed with the same setup. HD 115842 was observed soon
after the DY Cen exposures on all nights to correct for telluric lines in the DY
Cen spectrum. The S/N of the HD 115842 spectrum is 300 even for a single night's
observation.

In addition to these spectra from 2010, we draw on a spectrum obtained in 2003
from the 3.9m AAT with the University College London Echelle Spectrograph
(UCLES). The UCLES spectrograph was used with the 1.4" slit (slit length of
6")\footnote{The star was considered as a point source and standard
settings were used for the slit position.} and three individual exposures of
1200 s each were obtained. Wavelength coverage is from 4780 to 8800 \AA. The
final S/N in the summed spectrum is 38 to 40 in the continuum at 6300 \AA~to
6750 \AA~spectral region. Resolving power is about 40000 as estimated from
telluric [O\,{\sc i}] lines. Also, considered here is the 1989 spectrum used by
Rao et al. (1993). This was acquired at CTIO with the Blanco 4-meter telescope
and a Cassegrain echelle spectrograph. The slit width was 2"\footnote{No nebular emissions perpendicular to the dispersion were detected (Rao et al.
1993). Unfortunately, the information about the position angle of the slit is
not available.}, which gives a resolving power of about 18000 with the
wavelength coverage from 5480 to 6830 \AA\ in 1989. A 1992 spectrum with Blanco
telescope covered the interval  5480 to 7080 \AA\ at a resolving power of 35000
(see Giridhar, Rao \& Lambert 1996). 

DY Cen was identified as a RCB variable by Hoffleit (1930) from
well-determined minima in 1897, 1901, 1924, and 1929. No RCB type minima have
been recorded since 1958 (Bateson 1978, A. A. Henden 2010, private
communication). Thus, DY Cen is not an active RCB star and all the spectra
discussed here were taken when the star was at maximum light.

Quantitative interpretation of nebular emission lines requires knowledge of the
line fluxes and, hence, of the continuum fluxes across a spectrum. To obtain
such fluxes, we use UBVRI photometry of the star reported for the period 1989 -
2010. DY Cen has been slowly fading for the last 40 to 50 years (Bateson 1978;
Rao et al. 1993). This fading is independent of RCB-like events. We have
collected  UBVRI photometry reported in the literature. Figure 1 shows the
variation of B, V, and R from 1970 to 2010.

    \begin{figure}
\epsfxsize=8truecm
\epsffile{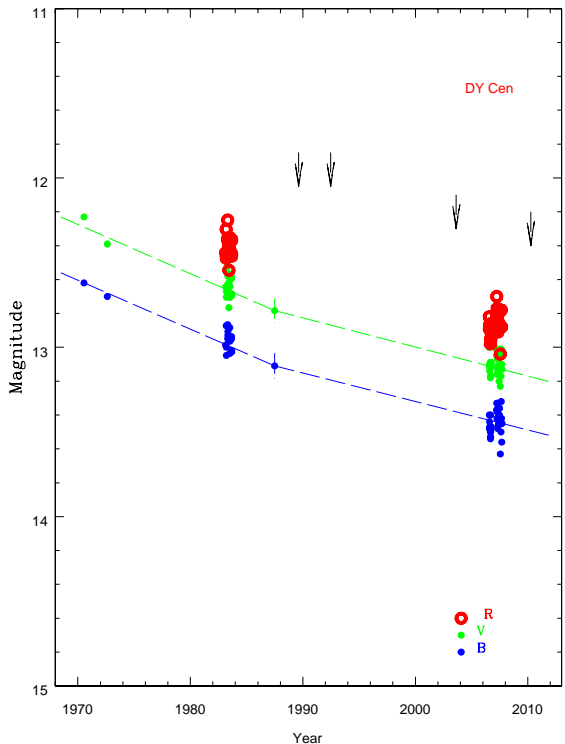}
\caption{Long term variations of light at maximum. The gradual fading
   of light from 1970 to 2010 of DY Cen in BV and R bands. The arrows
   indicate the dates of our spectroscopic
   observations mentioned in the text.  }

\end{figure}

The earliest published UBV magnitudes are from mid-1970
by  Marino \& Walker (1971):
V, B-V and U-B are 12.23, 0.39 and -0.62,
    respectively.
    U band measurements
    are few and were  obtained only in 1970  (see above) and during the 1982-83
    period at SAAO by Kilkenny et al. (1985).
     The 1972 measurements are from V.E. Sherwood (quoted by Rao, Giridhar \&
    Lambert 1993). Pollacco \& Hill (1991) obtained BV mesurements during 1987
    May and June. The 2007 BVRI measurements  are from the
    AAVSO. It is clear from the figure that there is a  gradual
    increase in magnitude over  40 years. The total variation is
    about 0.90 magnitude in V and  0.83  magnitude in B in 37 years:
  the (B-V) colour variations are
    minimal. Even (U-B) seems to be same in 1970 and 1982-83.

     Because of these very slow changes, we adopted the colours
     observed in 1987 for our spectroscopic observations in 1989 and the mid-2007
     photometry for our 2010 spectroscopic observations to estimate
     continuum fluxes.
The continuum fluxes are estimated from the UBVRI colours
after applying a correction for interstellar reddening for
an E(B-V) of 0.47 (Garc\'{\i}a-Hern\'{a}ndez et al. 2011b). Standard
relations:
      $A_{\rm V}$$\simeq $ 3.1 E(B-V) and E(B-V)/E(U-B) =0.72, $A_{\rm R}$/
     $A_{\rm V}$$\simeq $0.75 and $A_{\rm I}$/$A_{\rm V}$$\simeq $0.479 have been
     used (Cardelli et al 1989). Flux  calibrations given by Drilling \& Landolt (2000)
    are adopted.

Reddening-corrected line fluxes are given in the Table 1 for the 1989, 1992,
2003, and 2010 observations. The S/N of the 2010 spectrum is over 200 to 250.
The uncertainities in equivalent widths (EQWs) are 3\% in weaker lines (and less
in strong lines like H$_{\alpha}$). The photometric colours are more or less
unchanged so resulting errors in flux are also expected to be around 3\% or
better. Thus, the relative flux ratios are expected to be better than a percent.
The 2003 spectrum S/N is about 40 and the uncertainty in fluxes are around 5\%
level. The 1989 and 1992 entries are from Rao et al. (1993) and Giridhar
et al. (1996), respectively. For some lines, corrections for blending with
absorption or other emission lines are applied. In particular, the [S\,{\sc ii}]
6730.8 \AA\ line is  blended with the  C\,{\sc ii} 6731.1 \AA\ absorption line
on the redside, particularly on the 1989 July spectrum. A Gaussian fit to the
blueside of the forbidden line was used in estimating the absorption line
correction. Since the FWHM (after a minor correction instrumental broadening)
agrees with the FWHM  from 2010 spectrum, we suggest that  proper allowance is
made for the blending. In the 2010 spectra several C\,{\sc ii} lines are in
emission. The emission is seen only in the core of  absorption lines of the
multiplet of which 6731.1 \AA\ is a member (Figure 2). The contribution of this
line to the strong emission [S\,{\sc ii}] 6730.8 \AA\ is fortunately minor.
Thus, we trust the line ratio of [S\,{\sc ii}] 6716 \AA\ to  6731 \AA\ is not
unduly affected.

Gradual fading of DY Cen seems unlikely to be due to an increased presence of
dust along the line of sight to the star; a colour variation would likely result
from the dust. For a magnitude extinction the E(B-V) change of 0.32 is
expected, if the reddening law of the diffuse ISM is adopted. Even otherwise
RCBs in recovery phases typically show A$_{v}$/E(B-V) $<$ 4 - so one would
expect a change in E(B-V) of 0.25. Such colour changes are large and noticeable.
Any colour change $>$0.1 in E(B-V) is noticeable. We speculate that the fading
may be due to the star evolving to higher temperature at constant bolometric
luminosity, as expected for such stars. Jeffery \& Heber (1993) from a spectrum
obtained on 1987 April 17 derived an effective temperature of 19500 K. Our 2010
spectra suggest a spectroscopic ionization/excitation temperature of 25000 K.
Across the temperature range 19000 K to 25000 K, (B-V) is insensitive to
temperature: the change is only 0.07. If the bolometric magnitude is constant,
the V change over this temperature range is about 0.9 magnitude with V
increasing as the effective temperature increases. This is close to the observed
change in V (and B). The rate of evolution may be surprisingly rapid. 
Jeffery et al. (2001a) have monitored 12 EHe stars over a period of 10
years looking for the changes in their T$_{eff}$ and radii. The surface of four stars
was found to show heating (contraction) at rates between 20 and 120 degrees per
year in agreement with theoretical predictions based on models by Saio (1988).
DY Cen, which may be considered an EHe star, shows a heating rate
of 239 K degrees per year somewhat more rapid than EHe stars.

\begin{table*}
\centering
\begin{minipage}{180mm}
\caption{ \Large
    Dereddened Nebular line fluxes in DY Cen }

\small\begin{tabular}{lcllcrllllllll}
\hline\hline
     &  &    &   &   &   &   &   & & & &  &  &  \\
     &  &2010 &  &  & & &2003 (1992)$^{a}$& && 1989&  &  &  \\
\cline{3-6} 
\cline{8-9}
\cline{11-14}\\
   Ion &$\lambda c$ &Eq.w  & Flux$^{b}$ & (FWHM)${_0}$ & Rad.Vel&&Eq.w & Flux$^{b}$ && Eq.w  & Flux$^{b}$ & (FWHM)$_0$ & Rad.Vel \\
      & \AA\ &  (m\AA) &  x 10$^{-14}$ & km s$^{-1}$&km s$^{-1}$&&(m\AA)& x 10$^{-14}$ &  & (m\AA)  &x 10$^{-14}$ & km s$^{-1}$ &km s$^{-1}$\\
\hline
    &   &   &  &  &   &   & & &  &  &\\

    [O\,{\sc ii}] & 3726.0 & 433 & 10.2 & 34 &20.9& & & & &  &  &  &   \\

    [O\,{\sc ii}]& 3728.8 & 224 & 5.3 & 34 &18.8 & & & & & &   &         \\

    [N\,{\sc ii}] & 5754.6 & 38  & 0.28 & 41 &19.8& & & & &  &   &    &     \\

    [O\,{\sc i}]  & 6300.3 & 131 & 0.76 &28.5&20.2& &74 &0.46& &87 & 0.68 & 26 &23.3  \\

    [O\,{\sc i}] & 6363.8 &  52 & 0.29 &26.6&21.1& &33(67)&0.20(0.47)& &  &  & &   \\

    [N\,{\sc ii}] & 6548.0 & 1425& 7.25 & 41 &21.3& &478(46)&2.59(0.29)& &47 & 0.32 &  &25.8         \\

   $H \alpha$& 6562.8 & 9620& 48.6 &44.6& & &3798(2000)&20.4(12.4) &  &1473& 9.95 &    &         \\

    [S\,{\sc ii}] & 6716.5 & 350 & 1.61 & 34 &22.6& &157(67)&0.77(0.38) & &92 & 0.56 &34&22.2  \\

    [S\,{\sc ii}] & 6730.8 & 588 & 2.69 & 34 &22.0&  &169&0.82&  &83 & 0.50 & 34 & 22.4   \\

    [O\,{\sc ii}] & 7319.1 &  93 & 0.31 &43.7&    & & &    &  & &    &    &         \\

    [O\,{\sc ii}] & 7320.1 & 273 & 0.92 &46.6&    &  & &   &  & &    &    &         \\

    [O\,{\sc ii}] & 7329.7 & 154 & 0.52 & 45 &    &  & &   & & &     &    &         \\

    [O\,{\sc ii}] & 7330.8 & 157 & 0.53 & 45 &    & & &    & & &     &    &         \\

    [Fe\,{\sc ii}]& 4287.4 &  25 & 0.36 &10.4&21.1& & &  &  & & & &   \\

    [Fe\,{\sc ii}]& 4359.3 &  17 & 0.26 & 8.8&21.5& & &  &  & & & &     \\

    [Fe\,{\sc ii}]& 4413.8 &  13 & 0.20 & 8.6&    &  & &   &  & &  &   &   \\

    [Fe\,{\sc ii}]& 4243.9 &   7 & 0.10 &    &    &   & &   & & &  &  &   \\

\hline\hline
\end{tabular}

$^{a}$\ 1992 data from Giridhar et al. (1996) are given between bracketts\\
$^{b}$\ in ergs cm$^{-2}$ s$^{-1}$

\end{minipage}
\end{table*}

\begin{table*}
\centering
\begin{minipage}{160mm}
\caption{ \Large
    Flux ratios of H\,{\sc i} Paschen emission (nebular) lines in DY Cen in 2010 }

\small\begin{tabular}{lcllcrllll}
\hline\hline
     &  &    &   &   &   &   &   &   &  \\
     & & &  &  &  &  & &$T_{\rm e}$ 9425K& \\
     & & &  &  &  &  & &$n_{\rm e}$ 10$^{3}$ cm$^{-3}$& \\

   Line &$\lambda c$ &Eq.w  & Flux$^{a}$ & (FWHM)$_0$ &  Rad.Vel.    &   & j/j$_{\rm H \alpha}$(obs) & j/j$_{\rm H \alpha}$(Case B) &   \\
      & \AA\ &  (m\AA) &  x 10$^{-14}$ & km s$^{-1}$&  km s$^{-1}$ &  & &  &\\
\hline
    &   &   &  &  &   &   &   &  &\\

   p11            & 8862.8 & 162 & 0.34&34.7&20.4  &  & 0.0069 & 0.0059&        \\

   p12            & 8750.5 & 125 & 0.27&34.8&19.7 &  &0.0056 & 0.00489&       \\

   p14            & 8598.4 & 72 & 0.17 &36.7&20.2 &  &0.0034 & 0.0029 &        \\

   p17            & 8467.3 & 55 & 0.13 &34.4&20.1&   & 0.00275& 0.0015&        \\

   p20            & 8392.4 & 20 & 0.05 & 41 & &   &0.0010 & 0.00084 &     \\

\hline\hline
\end{tabular}

$^{a}$\ in ergs cm$^{-2}$ s$^{-1}$

\end{minipage}
\end{table*}

\section{Physical conditions in the nebula}

The suite of nebular lines provides  measures of the nebula's electron density,
temperature, and radius. Comparison of the spectra from 1989 to 2010 shows
evolution of physical conditions over these two decades. Comparison of the
lines' radial velocities with the binary's systemic velocity suggest that the
nebula may be symmetrically distributed about the binary.

    \begin{figure}
\epsfxsize=8truecm
\epsffile{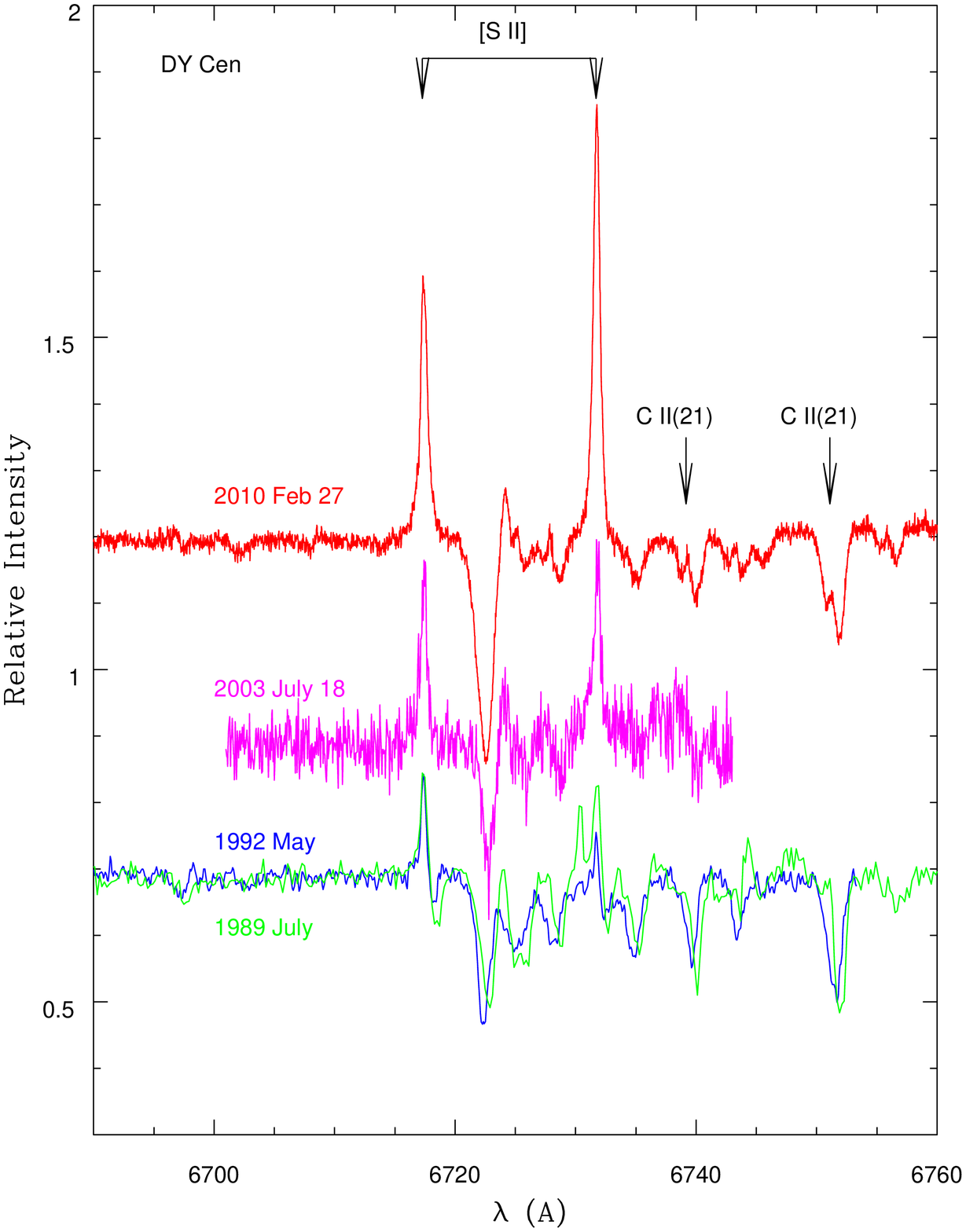}
\caption{Dramatic changes of the intensity ratio of [S\,{\sc ii}]
6716 \AA\ to 6731 \AA\ between 1989 and 2010.  From 1989 to 2003, the 6716 \AA\
line is stronger than or comparable in strength to the 6731 \AA\ line but
in the 2010 spectrum the 6731 \AA\ is obviously the stronger line.
    The apparent doubling of the 6731 \AA\ line in the 1989 spectrum may be due to
contamination by a [Fe\,{\sc ii}] line.}
\end{figure}

\subsection{Radial velocity}

In obtaining the nebula's radial velocity, we use a line's Ritz
wavelength as provided by the NIST
website.\footnote{http://physics.nist.gov/PhysRefData/asd.cfm}
Radial velocities from the individual lines in the 2010 spectra are
given in Table 1. There is no obvious change of velocity with wavelength
or ionic species and, therefore, we average these measurements to obtain
the mean velocity of 20.9$\pm{0.3}$ km s$^{-1}$  from  10
lines measured on each of the four  nights. From the 1989 and 2003 spectra,
   we obtain a radial velocity
of  22$\pm 2$  and  22.3$\pm 1.2$ km s$^{-1}$, respectively.
Thus, there is no indication of a velocity change from 1989 to 2010, although physical
conditions were evolving, i.e., the electron density increased
substantially  (see below) over the same period.

The sequence of 2010
VLT/UVES spectra shows very clearly the independence of the nebular
radial velocity from that of the  stellar photosphere.
The 2010 and other velocities led
Rao et al. (2012) to propose that DY Cen is a single-lined
spectroscopic binary.
Over four nights of the 2010 observations,
the photospheric velocity varies over a range of 22 km s$^{-1}$
but the nebular velocity is constant to within  1 km s$^{-1}$.
The overall mean velocity from the nebular lines is in good agreement
with the $\gamma$-velocity of 21.3$\pm0.5$ km s$^{-1}$
provided by the binary solution from the
photospheric absorption lines (Rao et al. 2012).
This close agreement in radial velocity suggests that the
nebula is symmetrically distributed about the binary. With the
estimate of the nebula's radius (see below) far exceeding the orbital
size, `nebula' seems an appropriate description for the region
emitting these forbidden lines.

\begin{figure}
\epsfxsize=8truecm
\epsffile{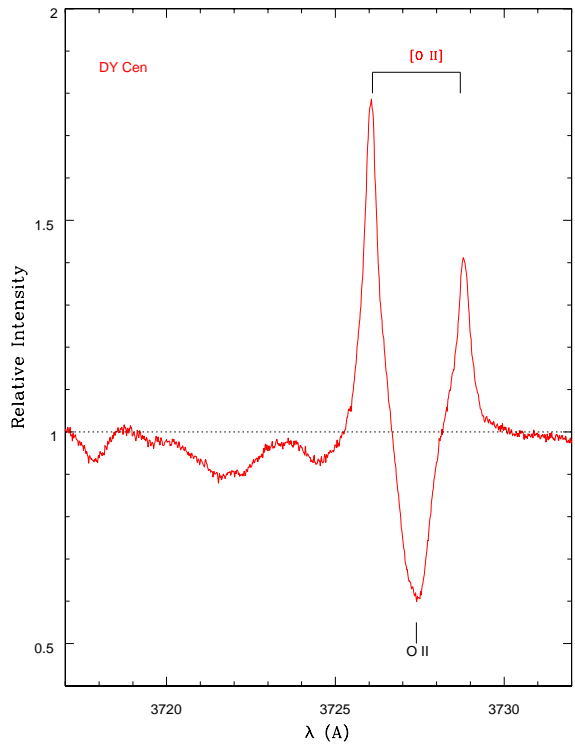}
\caption{The [O\,{\sc ii}] 3727 and 3729 \AA\  lines
   in the  2010 spectra. An absorption line of O\,{\sc ii} at 3727.3 \AA\
   affects the lines ratio slightly;  the estimated effect of this
line is to reduce the 3726/3729 line ratio by only a per cent.
}
\end{figure}

\subsubsection{Nebular line widths \& [Fe\,{\sc ii}] lines}

The nebular lines have full-widths at half maximum intensity (FWHMs) larger than
the instrumental width for the 1989 and 2010 spectra. The instrument-corrected
Gaussian FWHMs (FWHM)$_0$ of various lines are given in Table 1. There seems to
be a mild trend  for (FWHM)$_0$ of forbidden lines of N, O, and S to increase
with ionization potential. All widths are greater than the thermal widths. The
suggested expansion velocities are in the range 17 to 22 km s$^{-1}$.

Clearly, the [Fe\,{\sc ii}] lines have  much the  smallest
  width (comparable
   to the instrumental width) but share the radial velocity of
other forbidden lines.
  Several [Fe\,{\sc ii}] lines are present in the blue part
of the   2010 spectra  (Table 1). The spectrum of
    DY Cen obtained by Jeffery \& Heber (1993) on 1987 April 16 at a spectral
     resolving power of 27000 and displayed in
    Leuenhagen, Heber \& Jeffery (1994)  shows no trace of
[Fe\,{\sc ii}] emission lines. De Marco et al. (2002) obtained couple of
    blue spectra at a resolution of 35000 on 2001 April and September. The region
    around 4267 \AA\ displayed in their figure 12 shows  weak [Fe\,{\sc ii}] emission at
    4287 \AA\ is present but not as strongly as we see in our
    2010 spectra: the peak relative flux in 2001 was $\sim $ 1.07
whereas it is $\sim $ 1.13 in 2010. The radial velocities agree with
    those of other nebular lines (Table 1).
     The line widths obtained in 1989 spectrum
    agree with the ones determined from 2010 spectra.

    Usually [Fe\,{\sc ii}] lines
   arise from low excitation higher electron density regions
    (N$_{\rm e}$ $\sim $ 10$^{4}$- 10$^{6}$cm$^{-3}$; Perinotto et al. 1999, 
Viotti 1976). 
  Where from around the region around DY Cen
    could they arise? The [Fe\,{\sc ii}] lines share the radial
velocity of other forbidden lines which suggests that the region of formation
is linked to that of the general nebula. Then, might
  they be from the interface between the H\,{\sc ii}
    region and the surrounding neutral region at which the higher velocity
   expanding hot nebular gas  pushes into the neutral region?

    \begin{figure}
\epsfxsize=8truecm
\epsffile{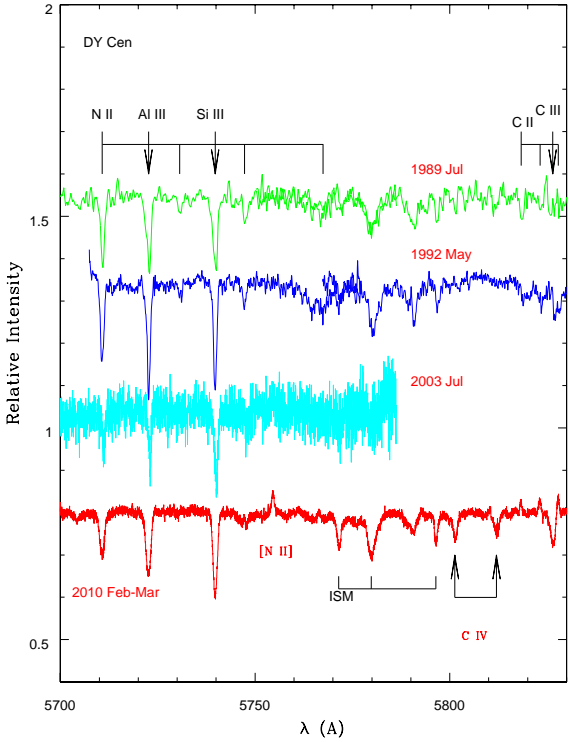}
\caption{ The wavelength region around 5750 \AA\ showing the appearance in
2010 of the  nebular line of [N\,{\sc ii}]
at 5755\AA.}
\end{figure}

\subsection{Electron density and temperature}

For the 2010 spectra, the electron density $N_e$ is obtainable from
two line ratios: [O\,{\sc ii}] 3726\AA/3729\AA, and [S\,{\sc ii}] 6716 \AA/6731\AA.
   For the 1989 and 2003 spectra with their more limited
spectral coverage, the [S\,{\sc ii}] ratio is the sole electron density
indicator.\footnote{Physical parameters
-- electron density and temperature --  were derived using the `PyNeb' software package
(Luridiana et al. 2012), using the atomic parameters listed in Garc{\'{\i}}a-Rojas, et al. (2012).}

From the 2010 spectra, the [O\,{\sc ii}] ratio gives $N_e = 3105\pm200$
cm$^{-3}$ and the [S\,{\sc ii}] gives $N_e = 3170\pm200$ cm$^{-3}$ and we adopt
the average $N_e = 3140$ cm$^{-3}$. In deriving these values we adopt the
electron temperature $T_e$ =  9350 K (see below). Figure 3 shows the [O\,{\sc
ii}] lines in the UVES spectrum.  The presence of O\,{\sc ii} absorption line at
3727.3 \AA\ between the two [O\,{\sc ii}] lines affects the estimation of the
[O\,{\sc ii}] line ratio slightly. Figure 2 shows the [S\,{\sc ii}] lines in the
1989, 1992, 2003, and 2010 spectra and the reversal from 1992 to 2010 in the
flux ratio of the lines. The [S\,{\sc ii}] ratio gives  $N_e = 620$ cm$^{-3}$
in  2003, $N_e \simeq 450$ cm$^{-3}$ in 1992,  and $N_e \simeq 290$ cm$^{-3}$ in
1989. Figure 4 displays the wavelength region around the [N\,{\sc ii}] at 5755
\AA\ in the 1989, 1992, 2003, and 2010 spectra, showing the appearance of 
[N\,{\sc ii}]5755\AA\ in 2010. In addition, the 6540-6590 \AA\ spectral region,
which covers H$_{\alpha}$, [N\,{\sc ii}] 6548 and 6583 \AA\ nebular lines, and
C\,{\sc ii} stellar wind lines, in 2003 and 2010 is compared in Figure 5.

For the 2010 spectra, the electron temperature is obtainable from the flux ratio
of the [N\,{\sc ii}] lines  (6583 + 6548)/5755. Because of blending of the 6583
\AA\ line with emission from a C\,{\sc ii} stellar wind line, we calculate its
strength from the 6548 \AA\ line and the electron density. From the [N\,{\sc
ii}] ratio, we obtain the temperature $T_e = 9350$ K. Note that our measured
upper limit to the [N\,{\sc ii}] 5755 \AA\ line in the 2003 spectrum (see Figure
4) provides an upper limit of 11000 K for T$_{e}$.

Jeffery \& Heber (1993) provide the composition of the star as He:H of
7.5:1 (or 11.52 to 10.65 in log scale). Our recent unpublished analysis of the
2010 spectrum is consistent with this number (11.52 to 10.76$\pm$0.20). After
helium next dominant gas in the atmosphere of DY Cen is hydrogen. If the nebula
has a similar composition to that of DY Cen's photosphere, the leading electron
donor is expected to be hydrogen despite the fact that, as a RCB star,  DY Cen
is H-poor. In the nebula, the electron density is set by the balance between
photoionization and radiative recombination. Photoionization by stellar
radiation is proportional to

\begin{equation}
N_HR_*^2\int \pi F_\nu\sigma_\nu d\nu/h\nu $=$ N_HR_*^2P(\nu) 
\end{equation}

where $N_H$ is the density of neutral hydrogen, $F_\nu$  the stellar flux,
$R_*$ the stellar radius,
$\sigma_\nu$ the photoionization cross-section and the integral is
evaluated from the Lyman limit to higher frequencies.
Radiative recombination occurs at a rate $N_e^2\alpha(T)$ where
$\alpha(T)$ is the temperature-dependent total radiative
recombination coefficient.

If the nebula was largely undisturbed
with a low degree of ionization (i.e., $N_e << N_H$ from 1989 to
2010, the electron density scaled as

\begin{equation}
\frac{(N_e^2\alpha(T))_{2010}}{(N_e^2\alpha(T))_{1989}} \propto
    \frac {R_*(2010)^2}{R_*(1989)^2} \frac {P(\nu)_{2010}}{P(\nu)_{1989}}
\end{equation}

The integrals $P(\nu)$ may be estimated from model atmospheres computed by
Jeffery et al. (2001b) for carbon-rich He stars: the ratio of the integrals is
about 42 in favour of the 25000 K model representative of the star in 2010
relative to the 19000 K model for 1989. This factor with the $R_*$ term assuming
evolution at constant luminosity leads to a predicted increase in electron
density by a factor of (42/2.25)$^{0.5}$ or 4 from 1989 to 2010. This is less
than the measured increase from the [S\,{\sc ii}] lines of a factor of 11
between 1989 and 2010 but is comparable to the factor of 5 increase between 2003
and 2010. This might suggest non-uniform increase of T$_{eff}$ of the star
during this period since we expect the bolometric luminosity to remain constant.
The rate of change of electron density between 1989 and 2003 is only a factor of
2 and from 2003 to 2010 is a factor of 5.

\begin{figure}
\epsfxsize=8truecm
\epsffile{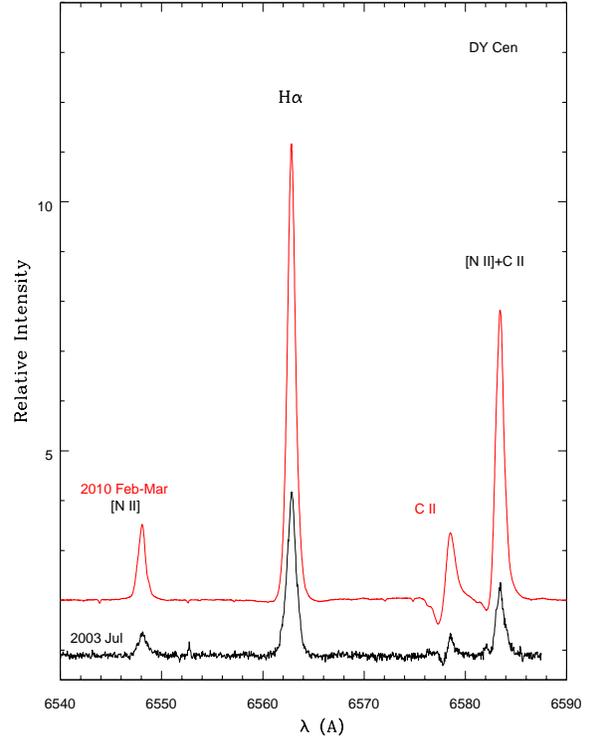}
\caption{The  H\,{\sc i} 6562 \AA, [N\,{\sc ii}] and C\,{\sc ii} lines
in the  2003 and 2010 spectra. }
   \end{figure}

\subsection{Ionic abundances}

Knowledge of $T_e$ and $N_e$ enables the ionic abundances to be calculated
(see Table 3). With the estimated $T_e$ and $N_e$ of 9350 K and 3140 cm$^{-3}$,
the 2010 line fluxes (Table 1) provide the O$^0$/H$^+$, O$^+$/H$^+$,
N$^+$/H$^+$, and S$^+$/H$^+$ ionic ratios listed in Table 3. The N$^+$/O$^+$
ratio is 0.46 and the S$^+$/O$^+$ is 0.019. Assuming O$^0$/O$^+$ = N$^0$/N$^+$ and
also S$^0$/S$^+$, and estimating the total O (= O$^0$ + O$^+$), N (=N$^0$ +
N$^+$), and S (=S$^0$ + S$^+$)  reduces the above ratios by about 15 \%.
   
\begin{table}
\centering
\begin{minipage}{60mm}
\caption{ \Large Derived ionic abundances}

\small\begin{tabular}{lcc}
\hline\hline
 & & (T$_{e}$=9350 K, N$_{e}$=3140 cm$^{-3}$)\\
 \hline
Ion  &   Lines  &  Abundance \\
\hline
N$^+$/H$^+$    & $\lambda$6548, $\lambda$5755 & 3.1 x 10$^{-5}$ \\
O$^0$/H$^+$    & $\lambda$6300, $\lambda$6363 & 1.2 x 10$^{-5}$ \\
O$^+$/H$^+$    & $\lambda$3726, $\lambda$3729 & 6.6 x 10$^{-5}$ \\
%               & $\lambda$7319, $\lambda$7330 &             \\
S$^+$/H$^+$    & $\lambda$6716, $\lambda$6731 & 1.3 x 10$^{-6}$ \\
\hline\hline
\end{tabular}
\end{minipage}
\end{table}
 
On the assumption that N/O $\simeq$ N$^+$/O$^+$, this nebular ratio is
about a factor of three larger than the photospheric
value which is N/O = 0.14 from Jeffery \& Heber's (1993) LTE analysis and
0.07 from our unpublished non-LTE analysis (in preparation).
Interestingly, the photospheric S/O ratio is 0.018 (Jeffery \& Heber 1993), a
value in good agreement with the nebula's ionic ratio. Thus, comparison of the
nebula and photospheric N/O/S ratios suggests that the former is N-rich
relative to the latter. This could be accounted for if the nebula represents
the envelope of  DY Cen ejected  after its first dredge-up when the envelope is
N-rich following the dredge-up of CN-cycled material. To a rough approximation,
much of the envelope at that point is N-rich owing to the conversion of
large amounts of the C to N, say N/O $\simeq$ 0.5. The He shell and deeper
layers of the star surviving the episode of envelope ejection may well
be far less N-rich because the N has been processed by $\alpha$-captures
to $^{22}$Ne.

\subsection{Hydrogen emission lines}

The 2010 spectrum shows Balmer and Paschen lines in emission. The Paschen lines,
    particularly the lower members  of the series, occur as photospheric
absorption lines on which nebular  emission is superposed.
We estimate emission line fluxes from higher members of the series where
absorption is minimal. Fluxes are given in Table 2.

Comparison between observed and predicted flux ratios is accomplished
using H\,{\sc i} recombination line ratios computed by Storey \& Hummer (1995)
for Case B - see Osterbrock \& Ferland (2006 - Table 4.4). We interpolate to
conditions $T_e = 9350$ K and $N_e = 1000$ cm$^{-3}$ for the flux ratio of
a Paschen line to H$\alpha$; the predictions are mildly sensitive to the adopted
$T_e$ and insensitive to $N_e$. Observed and predicted ratios for five clean
Paschen lines with respect to H$\alpha$ are given in Table 2. Good agreement
is seen suggesting that the lines including H$\alpha$ are optically thin.

    \begin{figure}
\epsfxsize=8truecm
\epsffile{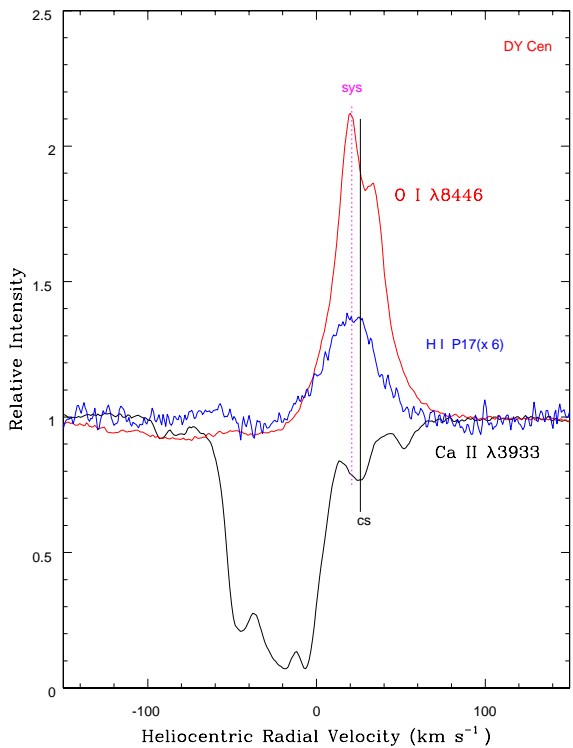}
\caption{Profiles of the O\,{\sc i} 8446 \AA\ (red) along with H\,{\sc i} Paschen 17 (six
   times magnified- blue) shown with Ca\,{\sc ii} 3933 \AA\ (black).
O\,{\sc i} profile is very different showing a superposed absorption (double peak)
   unlike
   other nebular lines, H$\alpha$ or Paschen lines. O\,{\sc i} 8446 \AA\ line is
produced by resonance fluorescence of H\,{\sc i} L$\beta$ with O\,{\sc i}
$\lambda$ 1025. Note the coincidence of the circumstellar absorption component
   in Ca\,{\sc ii} 3933 \AA\ with superposed absorption in O\,{\sc i} 8446 \AA.
   This absorption component is associated with the nebular and systemic
   radial velocity. }

\end{figure}

          The    O\,{\sc i} 8446 \AA\ line is present strongly in emission
	  (Figure 6) but
   presents a different profile than either H$\alpha$ or other nebular lines.
   Figure 6 shows a comparison with H\,{\sc i} Paschen 17 line. Since the FWHM of the
   O\,{\sc i} line is about the same as other nebular lines, it is
plausible to attribute the apparent doubling  to
superposed absorption rather than to a blend of two
   emission lines. O\,{\sc i}
   8446 \AA\ is produced by resonance fluorescence of H\,{\sc i} $L\beta$ with
   $\lambda$1025.76 of O\,{\sc i}. The excitation of  O\,{\sc i} by
   $\lambda$1025.76 is followed by emission at 11287 \AA\ and then by
emission at 8446 \AA.
    The O\,{\sc i}  exists in  either  the H$^{+}$ zone thanks to  charge exchange
   between oxygen and hydrogen or the neutral zone immediately surrounding the
   H\,{\sc ii} region.
The superposed absorption component in
   O\,{\sc i} 8446 \AA\ line implies that even H\,{\sc i} $L\beta$ might be
   experiencing the superposed absorption. It is a curious fact
   that a circumstellar absorption component at the same radial velocity
exists in Ca\,{\sc ii}
   $\lambda$3933  and is associated with stellar velocity (not the
   systemic velocity!). Although H$\alpha$ profile shows much wider radial
   velocity range it does not show any superposed absorption or asymmetry. It is
   centered  on the nebular velocity and has a range -42.5 to
   88.5 km s$^{-1}$ in radial velocity.

 The superposed absorption suggests the  presence of neutral gas  surrounding
the nebula. An estimate of the amount of this gas is made from the absorption
component of Ca\,{\sc ii} 3933 \AA~line (Figure 6). The equivalent width of this
component is about 55 m\AA. The expansion velocity from the FWHM  is estimated to
be about 8.5 km s$^{-1}$. Assuming it is optically thin, the column density of
Ca$^{+}$ is  6.4 x 10$^{11}$ cm$^{-2}$. The gf values from Morton (2003).
Assuming that Ca$^{+}$/H\,{\sc i} $\sim $ Ca/H  and further assuming solar
abundance ratio, the H\,{\sc i} column density estimated is 2.7 x 10$^{17}$
cm$^{-2}$. For the nebular radius estimated (in the next section) the mass of
neutral gas is estimated as 1.2 x 10$^{-5}$ $M_{\odot}$. This estimate is
dependent on  how the line of sight to the star intercepts the nebula.

\subsection{The size of the nebula}

De Marco et al. (2002) suggest that the distance to DY Cen is much more
than 4.8 kpc, as was suggested earlier by Giridhar, Rao \& Lambert (1996).
Assuming that DY Cen has the same M$_{v}$ (-3.0 mag) as its spectroscopic twin
the hot RCB star HV 2671 in the Large Magellanic Cloud (De Marco et al. 2002),
the estimated distance of DY Cen is about 7 kpc. In addition, the very nearby
B-type star HD 115842 (with E(B-V)=0.5 of ISM identical to DY Cen) may provide a
constraint that the distance should be greater than 5 kpc. HD 115842 seems to be
a hypergiant very similar to HD 169454 (M$_{v}$=-9.2 mag; Federman \& Lambert
1992) as indicated by the presence of Fe III emission lines in its spectrum. For
M$_{v}$=-9.2 mag, HD 115842 would be at a comparable distance of 5.4 kpc. Thus,
assuming a distance of 7 kpc for the star (De Marco et al. 2002) and the
H$\alpha$ flux given in Table 1 for 2010, the H$\alpha$ effective recombination
coefficient of 8.6 x 10$^{-14}$ cm$^{-3}$ s$^{-1}$ for the nebula's density and
temperature leads to an angular radius of the nebula of about 0.64 arcsec. Thus,
the slit width of 1.2 arcsec used for the 2010 observations should have covered
the nebular emission.

The geometrical radius corresponding to the angular radius is  6.5 x 10$^{16}$ cm.
The semi-major axis of the binary system is estimated to be 9.3/$\sin$
i$R_{\odot}$ (Rao et al. 2012)  where i is the angle of inclination of the binary
orbit to the line of sight. For $\sin$ i $\simeq$ 0.6, the semi-major axis is 1.1
x 10$^{12}$ cm. Thus, the nebular radius is $\sim$0.02 pc or 60000 times larger
than the binary orbit. The earlier estimated nebular expansion velocity (17 to 20
km s$^{-1}$), suggests an age of about 1100 years for the nebula. The ionized mass
 (mainly H which is providing electrons) within the nebula is about 3 x
10$^{-3}$ $M_{\odot}$.\footnote{The ionized H mass is estimated from the
expression (4$\pi/3 R^3$) x (n$_{p}$(=n$_{e}$)/m$_{p}$) where R is the estimated
radius of the nebula (=6.5 x 10$^{16}$ cm), n$_{p}$ is number of protons (= no.
electrons/cm$^{3}$) and m$_{p}$ is the proton mass.} Note that the main
electron donor is hydrogen because: i) no nebular emission lines of He I are
observed (i.e., no dominant nebular He I recombinations); and ii) the
photoionizations of H I dominate over He I by a factor of $\sim$35. Thus
most of the electrons are contributed by H I.

\section{Concluding remarks}

The nebula centred on  DY Cen, a hot RCB star, is surely intimately related to
the star because its radial velocity is coincident within about one km s$^{-1}$
with the systemic velocity of binary system to which DY Cen belongs.
The nebula's expansion velocity as estimated from the width of nebular forbidden
lines suggests that the nebula may have been created about 1000 years ago.

Spectra from 1989 to 2010 indicate that the star and the  nebula are
evolving. In apparent concert with an increase in the effective
temperature of the star, the electron density in the nebula
has increased several-fold over the two decades sampled by available
high-resolution spectra.

It is quite possible that the recent formation of the nebula, the changes in
nebular conditions over the preceding 20 years, and the increase in the
stellar effective temperature since 1987 and the steady fading of the star at
optical wavelengths are all manifestations of DY Cen's past and present
evolution. The nebula may be the product of an episode of common
envelope evolution that resulted in a primary (DY Cen) which became a
H-deficient supergiant. Such supergiants are expected to evolve rapidly
to a very hot luminous star at the tip of the white dwarf cooling track.

Interestingly, DY Cen shows a heating rate (contraction) more rapid than
EHe stars. In either of the two present evolutionary scenarios (white dwarf
mergers or a last helium flash) for the origin of EHe stars and RCBs, the star
has to become a cooler supergiant (for a second time) and starts contraction at
constant luminosity. Whether the rate of contraction is uniform in time or not
is not clear. In case of DY Cen this path of evolution might not be the one the
star passed through. The binarity of the star might introduce a different course
(see Rao et al. 2012 for some discussion). Further observations of the star are
most crucial and important in understanding the evolution of hydrogen deficient
stars.

One is hopeful that DY Cen's future behaviour will be followed
intensively spectroscopically and photometrically. In particular, it
will be important to separate behaviour linked to the 39 day orbital
period from the supposedly slower evolutionary trends as DY Cen
crossed the top of the H-R diagram.

\section{Acknowledgements}
We acknowledge the anonymous referee for useful comments that help to improve
the paper. We also acknowledge with thanks the variable star observations from
the AAVSO international database contributed by world wide observers used in
this research. This research has made use of the SIMBAD database, operated at
CDS, Strasbourg, France. Our sincere thanks are due to Simon Jeffery and Vincent
Woolf for supplying the AAT archival spectrum of DY Cen. N. K. R. would
like to thank David and Melody Lambert for their hospitality at UT, Austin where
part of this work is done. D.A.G.H.and A.M. acknowledge support for this work
provided by the Spanish Ministry of Economy and Competitiveness under grant
AYA-2011-27754.

\end{document}